\documentclass[aip,graphicx,reprint,twocolumn,amsmath,amssymb]{revtex4-2}

\draft % marks overfull lines with a black rule on the right
\usepackage{graphicx}
\usepackage{xcolor}
\usepackage{siunitx}
\usepackage[hidelinks]{hyperref}

\newcommand{\RPTU}{\affiliation{Fachbereich Physik and Landesforschungszentrum OPTIMAS, Rheinland-Pf\"alzische Technische Universit\"at Kaiserslautern-Landau, 67663 Kaiserslautern, Germany}}
\newcommand{\Fraunhofer}{\affiliation{Fraunhofer Institute for Industrial Mathematics ITWM, Fraunhofer-Platz 1, 67663 Kaiserslautern}}

\begin{document}

%\preprint{}

\title{Rapid-prototyping of microscopic thermal landscapes in Brillouin light scattering spectroscopy} %Title of paper



\author{Matthias R. Schweizer}
\email{mschweiz@rptu.de}
%\thanks{}

\RPTU

\author{Franziska Kühn}
\RPTU

\author{Malte Koster}
\RPTU

\author{Georg von Freymann}
\RPTU
\Fraunhofer

\author{Burkard Hillebrands}
\RPTU

\author{Alexander A. Serga}
\email{serha@rptu.de}
\RPTU


\date{\today}

\begin{abstract}
Since temperature and its spatial and temporal variations affect a wide range of physical properties of material systems, they can be used to create reconfigurable spatial structures of various types in physical and biological objects. This paper presents an experimental optical setup for creating tunable two-dimensional temperature patterns on a micrometer scale. As an example of its practical application, we have produced temperature-induced magnetization landscapes in ferrimagnetic yttrium iron garnet films and investigated them using micro-focused Brillouin light scattering spectroscopy. It is shown that, due to the temperature dependence of the magnon spectrum, temperature changes can be visualized even for microscale thermal patterns.
\end{abstract}

\pacs{}% insert suggested PACS numbers in braces on next line

\maketitle %\maketitle must follow title, authors, abstract and \pacs

\section{Introduction}
Temperature is crucial in many scientific fields, including biology, materials science, physics, and chemistry.
By changing the temperature, it is possible to achieve both irreversible changes of the medium or to implement non-destructive reversible modifications in the system parameters, as it indirectly determines such internal quantities as saturation magnetization \cite{Vogel2015, Vogel2018, Bozhko2019}, dielectric permittivity, resistivity, refractive index, and mechanical expansion coefficients, among others.\cite{Manganelli2020, FreitasNeto2012} Gradients in temperature induce transport processes, and, thus, by adjusting their spatial distribution, it is possible to tune those transport processes.

In this regard, an important task is to create specific microscopic thermal profiles in samples of different types, which can then be altered in space and time on demand. \cite{Bozhko2019, Schneider2020, Schweizer2022, Agrawal2014, Durdevic2019, Hamada2018, Wang2020, Hinzke2011, Martens2018}
This task occurs not only in basic research. It also arises, for instance, in the development of thermoelectric and spin-caloritronic devices \cite{SantaClaraGomes2019} that utilize thermal energy or for prototyping functional devices that use a heterogeneous medium to process information carried by waves passing through that medium \cite{Kolokoltsev2012, Dzyapko2016, Davies2015a, Vogel2015, Vogel2018, Weiler2012}.
The structure required for the latter task can be introduced into the system under study in advance using sophisticated fabrication methods. \cite{Schneider2020, Heinz2020, Obry2013, Goncalves2021} Although the relevant methods have been continuously revised and improved over decades, they always involve considerable effort and they may irreversibly alter the samples used. A different approach for the fabrication of desired topographies is to use reversible parameters controlled by reconfigurable temperature landscapes. \cite{Schweizer2022, Vogel2015, Vogel2018, Durdevic2019, Vogel2020}

In this work, we present an experimental setup which can generate complex temperature landscapes with resolution down to the micrometer scale by projecting optical intensity patterns onto the surface of a sample. These intensity patterns are created by means of a wavefront modulation technique (see Sec.~\ref{sec:SLM}). 
Moreover, we have combined this technique with Brillouin light scattering (BLS) spectroscopy (see Sec. ~\ref{sec:BLS}) to enable measuring the effect of temperature inhomogeneities on different types of wave excitations (Sec.~\ref{sec:experiment}).  
BLS spectroscopy is widely used in various fields \cite{Meng2016, Palombo2019, Kwon2013, Sandweg2010, Bozhko2020, Yoshida2011}, including condensed matter physics, materials science, and biophysics, to understand the physical properties of samples at the microscopic level. \cite{Sebastian2015}
Specifically, we applied the developed all-optical setup to manipulate and measure the characteristics of magnons, i.e., collective spin excitations, in ferrimagnetic films of yttrium iron garnet (YIG, $\mathrm{Y}_3\mathrm{Fe}_5\mathrm{O}_{12}$).
Although this work is focused on magnons, the demonstrated approach can also be applied to phonon systems as well as to a variety of biological and material science applications if the behavior of the systems under study relies on temperature-dependent material properties.

\section{Experimental setup}
\begin{figure}[t]
\includegraphics{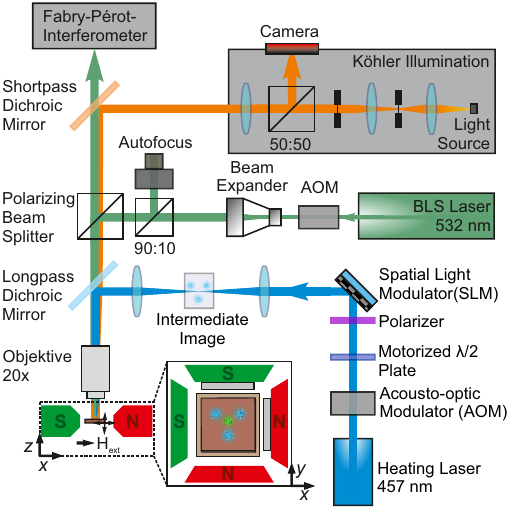}
\caption{\label{fig:setup} Schematic view of the experimental setup. The imaging beam path (orange) generates a magnified image on a CCD-camera. The BLS module (green beam path) serves as the primary means of detection, while the heating module (blue beam path) is used for the creation of the temperature landscapes under investigation.}
\end{figure}

Figure~\ref{fig:setup} shows a schematic view of the presented experimental setup.
It can be divided into three modules: (i) the heating module, which projects optical intensity patterns onto the sample surface to form a desired temperature landscape; (ii) the detection module for the spectroscopic observation of magnons or phonons; and (iii) the imaging module to align the sample to the probing beam of the detection module. In the following section, these modules will be described individually.
In addition, the setup is equipped with a 2D in-plane vector magnet to allow for different magnetization regimes.

The control of these modules has been implemented by using the automation framework \textit{thaTEC:OS} (THATec Innovation GmbH), while the following data evaluation was performed using \textit{Python}-based software libraries such as \textit{PyThat} and \textit{xarray}. \cite{THATec, PyThat, Hoyer2017, Virtanen2020}

\subsection{Heating module for the creation of microscopic 2D temperature landscapes}
\label{sec:SLM}
The developed heating module creates temperature landscapes employing optical holograms and Fourier optics \cite{Vogel2015, Vogel2018, Schweizer2022, Durdevic2019}. The holograms are formed by the utilization of phase-based wavefront modulation: A spatial light modulator (SLM), i.e., a liquid-crystal array on a silicon chip, imprints a spatial distribution of phase shifts onto an incoming planar wavefront. The corresponding phase map for a given desired intensity distribution can be calculated by means of the Gerchberg-Saxton algorithm for iterative phase retrieval. \cite{Gerchberg1972, Alsaka2018} 
By placing a lens at its focal distance away from the SLM, an intermediate image of the resulting intensity pattern is created \cite{goodman2005introduction}. By adding an iris aperture in this optical plane, the specular reflex---the part of the light which is unaffected by the phase shift of the SLM---as well as higher orders of the diffraction pattern are blocked. Due to the linear properties of Fourier optics, the aforementioned phase maps can be easily adjusted by adding Zernike polynomials, a class of orthogonal functions that describe the wavefront aberrations of rays \cite{Voelz2011, Yen2021, Zernike1934}. In particular, the ``tilt''-terms can be used to shift the position of a given intensity landscape across the sample without the necessity for any movable mechanical parts\cite{Yen2021, Voelz2011, goodman2005introduction}.

The combination of another lens and the microscope objective serves as a microscope configuration which projects the intermediate image onto the sample. The numerical aperture of \hbox{NA=0.45} results in spot sizes of about \SI{2}{\micro\meter}.

In our experiment, we use the SLM \textit{Hamamatsu X10468} with a resolution of \hbox{792x600 pixels} and a pixel pitch of \SI{20}{\micro\meter} together with the laser \textit{Cobolt Twist} emitting light at \SI{457}{\nano\meter}. The wavelength has been chosen since it is known to exhibit strong absorption for YIG \cite{Doormann1984}. Before targeting the SLM, the beam is expanded to a diameter of approximately \SI{1}{\centi\meter}, therefore using a large fraction of the SLM's aperture.

This combination of small focal spot size and strong absorption results in a high density of absorbed optical power and imposes the need for additional precise control of the total heating power.
While it is possible to change the laser's output power via the driving current, this bears the risk of mode shifts, making it unsuitable for automated parameter sweeps.
However, the control of optical power can be achieved by a simple sequence of a half-wave plate and a linear polarizer, where the rotation of the half-wave plate can change the polarization axis, while the linear polarizer will transmit only the polarization component required by the SLM. Due to the large beam diameter, the absorption does not damage the polarizer. By calibrating the motorized waveplate to the characteristic transmission proportional to $\cos^2(2 \alpha)$, where $\alpha$ is the relative angle between the transmission axis of the polarizer and the fast axis of the half-waveplate, a linear power control can be achieved.

When working with thermal landscapes (especially on microscopic length scales), thermal diffusion cannot be avoided. The transport of heat along the temperature gradient, i.e., thermal diffusion, results in a blurred temperature distribution. In thermal equilibrium, the resulting size of a hot spot would depend on the thermal conductivity of the material, its heat capacity, and the distribution of heat influx. However, this effect can be somewhat softened by inducing heat in a pulsed microsecond regime. In this way, a measurement can be taken while a small volume is hot, but before the heat spreads into adjacent material. After each measurement, the system thermalizes to a homogeneous temperature state which can once again be heated quickly.
We add the ability to chop the laser beam by adding an acousto-optic modulator (AOM) \textit{G\&H AOM 3080-125} to the heating module, where the first diffraction order is used, which leads to negligible transmission in off-state. Since it provides rise- and fall-times below \SI{100}{\nano\second}, this is also a suitable tool to measure the dynamic effects of a rapid temperature increase or decrease. However, due to the nature of thermal transport, the effective times for temperature change are highly dependent on the sample under investigation and may be significantly longer.

Consequently, the presented heating module offers the means to create complex reconfigurable intensity landscapes with rise and fall times below \SI{100}{\nano\second} and the ability to continuously change the applied power in the range from \SIrange{0}{50}{\milli\watt}. The combination of both AOM and external power control offers the possibility of constant average power, even for altered duty cycles of the pulsed heating. 

\subsection{Magnon detection via micro-focused Brillouin light scattering spectroscopy}
\label{sec:BLS}
In this experiment, the density of magnons is measured by means of micro-focused BLS spectroscopy with sub-micrometer spatial resolution \cite{Sebastian2015}. This technique facilitates the frequency shift of light due to the inelastic scattering on collective excitations such as magnons. In this way, not only the overall density of particles can be detected but also the corresponding frequency, which contains further information about the character of an investigated system. While this technique can be used for phonons or magnons alike, in this work, we will focus on the detection of magnons.

The probing light source is a \textit{Coherent Verdi} laser operating at a wavelength of \SI{532}{\nano\meter}. It provides a narrow spectral linewidth and stable output power, but also has several side modes up to \SI{7}{\giga\hertz}. In order to suppress these side modes, the beam passes a polarizing beam splitter. Since scattering with magnons causes a change of the polarization\cite{Wettling1975, Geilen2022}, the inelastically scattered component is now transmitted through the polarizing beam splitter towards the interferometer
and the side modes are blocked, thereby improving the signal-to-noise ratio considerably.
The elastically scattered portion of the probing beam maintains its polarization and is used for an autofocus system to stabilize the distance between the objective and the sample.

\begin{figure}[tb]
\includegraphics{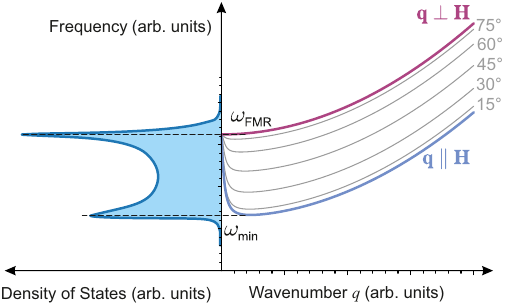}
\caption{\label{fig:theory} Schematic representation of the dispersion curves calculated for different propagation angles of the lowest volume spin-wave mode in a tangentially magnetized film of yttrium iron garnet with a thickness of several micrometers (right) and the corresponding density of magnon states (left).} \looseness=-1
\end{figure}

Both BLS beam and the SLM image (see Sec.~\ref{sec:SLM}) are projected onto the sample through the same microscope objective.
As a result, both spot sizes produced are within the range of approximately \SI{2}{\micro\meter}. This also implies that the system is capable of detecting waves with wavelengths as short as \SI{4}{\micro\meter}. Essentially, the high spatial resolution of the probing beam comes with the added advantage of being able to detect a wide range of wavenumbers simultaneously.
Although it is impossible to resolve a single state at one given wavenumber, this approach allows to capture a broad ensemble of spin waves (see, e.g., Fig.~\ref{fig:theory}) with a single measurement. 
In this way, it is possible to detect magnon occupation at wavenumbers from near zero to more than
\,\SI{5}{\radian\per\micro\meter}.

\begin{figure*}[tb]
\includegraphics[width=\textwidth]{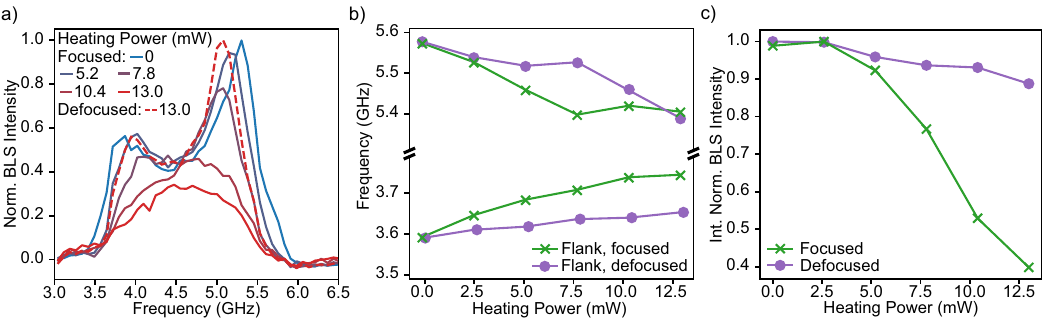}
\caption{\label{fig:power_dependence} Influence of the laser-induced temperature increase. Panel a) shows several magnon spectra measured at increasing heating powers. Both a shift of the high frequency edge of the spectrum towards lower frequencies and an overall decrease of BLS signal can be observed. The dashed line shows a spectrum which was measured with an intentionally defocused heating laser spot. Panel b) shows a quantitative relation between heating power and the magnitude of the corresponding frequency shift, whereas panel c) depicts the simultaneous decrease of integrated BLS intensity in the frequency range from \SIrange{3.5}{6}{\giga\hertz}. \looseness=-1}
\end{figure*}

In general, the probing laser, like the heating laser, can transfer heat into the system, which would distort the results. Therefore, the probe laser power has been chosen in such a way that it does not have an effect on the thermal spectrum of the YIG sample. In order to further reduce its influence, another \hbox{AOM \textit{G\&H 3080-195}} is integrated in the measurement beam path. In this way, it is possible to chop the measurement beam in order to concentrate the necessary optical intensity on the window of interest, while avoiding unnecessary heat transfer between measurement cycles. Furthermore, in the case of YIG, the optical absorption for \SI{532}{\nano\meter} is only about \SI{10}{\percent} of that at \SI{457}{\nano\meter} \cite{Doormann1984}.

\subsection{Imaging system}
While the scope of this article is on purely thermally induced landscapes, it should be mentioned that the majority of experiments will require the ability to align a given intensity pattern to a previously patterned structure.
Accordingly, a conventional microscopy illumination module has been implemented, as shown by the orange beam path in Fig.~\ref{fig:setup}. Due to the illumination with wavelengths longer than \SI{550}{\nano\meter}, the illumination is mostly separated from both the probing beam path and the heating beam.
However, as a result of the non-negligible reflectivity of the integrated dichroic beamsplitter, both the probing and heating beams can be imaged with high sensitivity by the \textit{Thorlabs CS165MU1} CCD camera.
This module enables the integration of additional electrical manipulation tools such as microstrip microwave spin-wave antennas, interdigital acoustic transducers, or contacts for DC measurements. 

\section{Exemplary Measurements}
The following section presents various aspects of the developed setup using exemplary measurements of the thermal magnon spectrum of an in-plane magnetized YIG film with a thickness of \SI{2.1}{\micro\meter}.
This spectrum is depicted in Fig.~\ref{fig:theory} for the lowest magnon mode.
Having a quasi-uniform profile over the film thickness, this mode exhibits the largest scattering cross section compared to the higher thickness modes and, thus, makes a dominant contribution to the recorded BLS signal. \cite{Bozhko2020} The experimental spectrum measured without heating (see Fig.~\ref{fig:power_dependence}a) agrees reasonably well with the density of states derived \cite{Rezende2020} from the theoretical spectrum in Fig.~\ref{fig:theory}. The shown density of states has two maxima: one around the frequency of ferromagnetic resonance (FMR) and one at the lowest frequency of the spectrum. These maxima correspond to the flattest parts of the dispersion curves, where magnons with different wavenumbers have close frequencies. As described in Sec.~\ref{sec:BLS}, all these magnons can be detected simultaneously due to the large wavenumber range of the micro-focused BLS module.
\label{sec:experiment}

\subsection{Changing the heating power in a single spot}
\label{sec:power_dependence}
The straightforward approach to determine the degree to which heating affects the thermal spectrum of a YIG film is to vary the applied total laser power while both the probing spot and the heating spot are aligned.
The results of a corresponding measurement are shown in Fig.~\ref{fig:power_dependence}.

The first important observation in Fig.~\ref{fig:power_dependence}a is the qualitative deformation of the BLS intensity spectrum as the heating power is increased. Most notably, both the high- and low-frequency peaks shift towards the center of the spectrum (Fig.~\ref{fig:power_dependence}b, purple lines).
The down-shift of the high-frequency peak is in good agreement with a change of the FMR frequency $\omega_\mathrm{FMR} = \mu_0 \gamma \sqrt{H_\mathrm{i}(H_\mathrm{i}+M_\mathrm{s}(T))}$, where $\gamma$ is the gyromagnetic ratio, $M_\mathrm{s}(T)$ is temperature-dependent saturation magnetization, and $H_\mathrm{i}$ is an internal magnetic field, which, for an in-plane magnetized film, is equal to the externally applied field $H_0$. In this case, the increasing temperature reduces $M_\mathrm{s}$, which in turn causes a shift towards lower frequencies.

In contrast, for the low-frequency peak at the bottom of the spectrum $\omega_\mathrm{min} \approx \mu_0 \gamma H_\mathrm{i}$, the observed frequency shift cannot be explained by the same effect, simply because $\omega_\mathrm{min}$  is expected to respond much more subtly to the change in the saturation magnetization. 

The reason for the observed upward frequency shift may be a non-uniform magnetization of the heated sample, which leads to an increase in the magnetic field within a small heated area of the sample, surrounded by an undisturbed magnetic medium with higher magnetization. As a result, $H_\mathrm{i}$ exceeds $H_\mathrm{0}$ and $\omega_\mathrm{min}$ increases.

For the highest observed heating powers, the overall shape of the spectrum begins to change. The previously clearly visible peaks at the edges of the spectrum gradually disappear. The whole spectrum slowly morphs into a very broad peak.
A possible mechanism behind this observation is that the perturbation of the structure itself is on the scale of the wavelength of the observed magnons. 

\begin{figure*}[tbh]
\includegraphics{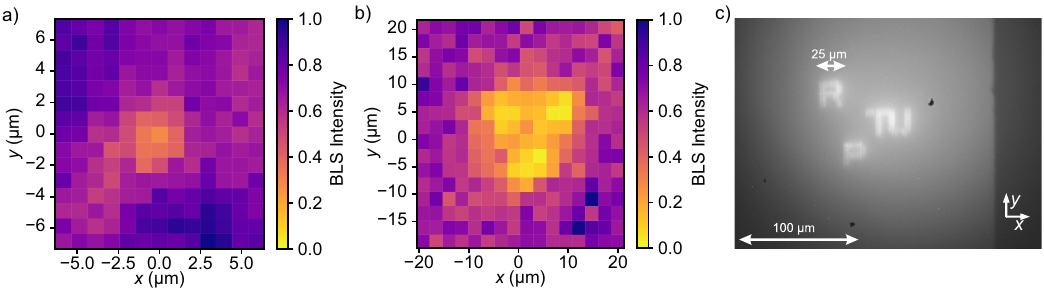}
\caption{\label{fig:single_spot} \label{fig:RPTU} \label{fig:triangle} Demonstration of the lateral resolution and complexity of optical intensity/thermal landscapes. a) 2D scan of the frequency-integrated BLS intensity over a single heated spot. b) 2D scan of three spots in a triangular configuration. The distance between the spots is \SI{13}{\micro\meter}. c) Intensity pattern showing the logo of the Rheinland-Pf\"alzische Technische Universit\"at Kaiserslautern-Landau. The image was obtained by projecting the reflection of the intensity landscape onto the CCD camera. The shape consists of many small spots separated by a distance of approximately \SI{2}{\micro\meter}.}
\end{figure*}

Now, in this particular setup, the heating beam can easily be defocused by adding a focus/defocus ($Z_2^0$) term of the Zernike polynomials to the SLM phase map (see sec~\ref{sec:SLM}).\cite{Yen2021} This slight defocus smoothens the intensity gradients at the edges of the heating spot, thereby creating a more homogeneous temperature profile. 
%At this point, it should be mentioned, that while the peak heating power \textit{density} is thereby reduced, the \textit{total} heating power remains constant. 
Using this technique, the influence of small feature sizes can be readily suppressed.
As a result, it can be seen from the red dashed curve in Fig.~\ref{fig:power_dependence}a and the lower purple line in Fig.~\ref{fig:power_dependence}b that the low-frequency peak shift is now significantly reduced, which indicates that this is indeed a feature of localized heating. 
%This interpretation is supported by the work of Beaulieu \etal suggesting that the magnon linewidth in YIG actually decreases with temperature.\cite{Beaulieu2018}

Another important observation is the overall decrease in BLS intensity 
(Fig.~\ref{fig:power_dependence}c). However, heating can only increase the quantity of thermally excited magnons. Thus, the decrease in BLS intensity should be associated with a decrease in BLS sensitivity reported by Olsson \textit{et al.}\cite{Olsson2018, Olsson2017} It is notable that while Olsson \textit{et al.} observed an intensity drop of \SI{25}{\percent} over \SI{60}{\kelvin}, we see a much stronger decrease, suggesting an even stronger temperature change.

\subsection{Lateral scan of intensity landscapes}
One of the advantages of reconfigurable structures is the ability to perform lateral measurements by moving the temperature landscape instead of the probing beam. For instance, in the case of constraints such as prefabricated structures, it is possible to fix the probing spot in a constant position. In this way, the change in the temperature landscape is the only factor that needs to be taken into account in the experiment. This concept holds true even for undesired features, such as impurities and surface defects, whose contribution to the BLS signals can be suppressed completely.

Figure~\ref{fig:single_spot}a shows the false-color representation of the frequency-integrated BLS intensity as a function of the position of the heating spot. As already seen in Sec.~\ref{sec:power_dependence}, the integrated BLS intensity drops significantly when both the heating and probing laser are aligned. The decrease in BLS intensity indicates a high temperature. Around this point, the intensity increases continuously until it reaches the ambient level. This behavior indicates that the temperature also increases continuously at greater distances from the probing point. The observed spot size of approximately \SI{2}{\micro\meter} agrees well with the estimated size of the probing and heating laser spots.

The same type of 2D scan is shown in Fig.~\ref{fig:triangle}b, but in this case, a two-dimensional pattern of three spots arranged in the shape of an equilateral triangle with a side length of \SI{13}{\micro\meter} is depicted.
Despite the small distance between the points, a rather good contrast can be achieved, which enables a clear visual separation of the hot spots. However,  between the spots, the BLS intensity does not reach the ambient level. This behavior can be attributed to thermal diffusion, which is to be expected in this type of measurement. In fact, the effect of thermal diffusion can also be seen in the form of a blurred BLS intensity distribution around all three spots.

The possible complexity of the patterns which can be achieved with the presented method is illustrated by a CCD image of an intensity pattern, projected on the surface of a sample, as it is shown in Fig.~\ref{fig:RPTU}c.
The pattern consists of a set of spots spaced at a distance of \SI{3}{\micro\meter} in the shape of a logo with a total size of \SI{100}{\micro\meter}. This image illustrates the power of this technique: While it is well suited for creating small features, it can also be used to create structures with form factors nearly two orders of magnitude larger.

\section{Conclusion}
% We present an experimental setup for the measurement of artificial reconfigurable 2D temperature landscapes. 
% The investigation of gaseous magnons under local heating reveals a qualitative change of its spectrum, when the temperature is strongly increased in a micro-sized area. This effect does not occur when the same amount of heat is distributed in a slightly larger area.
% Furthermore we have shown a lateral characterization of temperature landscapes, which reveals strong thermal contrast on the micro-meter scale. While the scope of this work focused on comparably simply model structres, much more elaborate structures can be created with this technique.
We present an all-optical experimental setup for direct BLS observations of quasiparticles such as magnons in reconfigurable microscopic thermal landscapes. This setup provides the ability to quickly create arbitrary landscapes for scientific measurements, as well as rapid prototyping, which can significantly reduce the time and opportunity cost in the development and measurement cycle. 

As a proof of concept, the study of gaseous magnons under localized heating reveals a qualitative change in their spectrum when the temperature is strongly increased in the microscopic region. This effect is not observed when the same amount of heat is distributed over a slightly larger area. Furthermore, we have shown the spatial characterization of temperature landscapes that reveals a strong thermal contrast at the micrometer scale. Although we have focused on relatively simple model structures in this work, much more complex temperature profiles can be created using this heating technique.

\begin{acknowledgments}
This study was funded by the Deutsche Forschungsgemeinschaft (DFG, German Research Foundation)---TRR 173/2---268565370 Spin+X (Project B04).
\end{acknowledgments}

\bibliography{abbr_2D_landscapes}

\end{document}